\documentclass{aa}
\makeatletter
\renewcommand*\aa@pageof{, page \thepage{} of \pageref*{LastPage}}
\makeatother

\usepackage{natbib}
\bibpunct{(}{)}{;}{a}{}{,} % to follow the A&A style
\usepackage{amssymb}
\usepackage{url}
\usepackage{hyperref}
\usepackage{graphicx}
\usepackage{longtable}
\usepackage[usenames,dvipsnames]{color}
\usepackage{booktabs}
\usepackage{lscape,xcolor}

\usepackage{multirow}
\usepackage{newtxtext}
\usepackage{amsmath}
\usepackage{enumitem}
\usepackage[T1]{fontenc} %to get ł

\def\Msun{\hbox{$\rm\thinspace M_{\odot}$}}

\begin{document}

 \title{%ISM removal in passive galaxies II: 
 Charting the main sequence of star-forming galaxies out to redshifts $z\lesssim5.7$
 %\thanks{Reduced images as FITS files are only available at the CDS via anonymous ftp to \protect\url{cdsarc.u-strasbg.fr (130.79.128.5)} or via
%\protect\url{http://cdsarc.u-strasbg.fr/viz-bin/qcat?J/A+A/562/A70}}
 }
 
\titlerunning{Main sequence of star-forming galaxies out to $z\lesssim5.7$}
\authorrunning{Koprowski et al.}

\author{M.~P.~Koprowski\inst{\ref{inst:tor}}
\and
J.~V.~Wijesekera\inst{\ref{inst:tor}}
\and
J.~S.~Dunlop\inst{\ref{inst:roe}}
\and
D.~J.~McLeod\inst{\ref{inst:roe}}
\and
M.~J.~Micha{\l}owski\inst{\ref{inst:poz},\ref{inst:roe}}
\and
K.~Lisiecki\inst{\ref{inst:tor},\ref{inst:ncbj}}
\and
R.~J.~McLure\inst{\ref{inst:roe}}
        }

\institute{
Institute of Astronomy, Faculty of Physics, Astronomy and Informatics, Nicolaus Copernicus University, Grudzi\c{a}dzka 5, 87-100 Toru\'{n}, Poland, {\tt drelkopi@gmail.com}\label{inst:tor}
\and
Institute for Astronomy, University of Edinburgh, Royal Observatory, Edinburgh EH9 3HJ, UK \label{inst:roe}
\and
Astronomical Observatory Institute, Faculty of Physics, Adam Mickiewicz University, ul.~S{\l}oneczna 36, 60-286 Pozna{\'n}, Poland \label{inst:poz}
\and
National Centre for Nuclear Research, Pasteura 7, 093, Warsaw, Poland \label{inst:ncbj}
}
\abstract{
We present a new determination of the star-forming main sequence (MS), obtained through stacking 100k $K$-band-selected galaxies in the far-infrared (FIR) {\it Herschel} and {\it James Clerk Maxwell} Telescope (JCMT) imaging. By fitting the dust emission curve to the stacked FIR photometry, we derive the IR luminosities ($L_{\rm IR}$), and hence the star formation rates (SFRs) out to $z\lesssim 5.7$. The functional form of the MS is found, with the linear SFR-$M_\ast$ relation that flattens at high stellar masses and the normalization that increases exponentially with redshift. We derive the corresponding redshift evolution of the specific star formation rate (sSFR) and compare our findings with the recent literature. We find our MS to be exhibiting slightly lower normalization at $z\lesssim 2$ and to flatten at somewhat larger stellar masses at high redshifts. By deriving the relationship between the peak dust temperature ($T_{\rm d}$) and redshift, where $T_{\rm d}$ increases linearly from $\sim 20$\,K at $z=0.5$ to $\sim 50$\,K at $z=5$, we conclude that the apparent inconsistencies in the shapes of the MS are most likely caused by the different dust temperatures assumed when deriving SFRs in the absence of FIR data. Finally, we investigate the derived shape of the star-forming MS by simulating the time evolution of the observed galaxy stellar mass function (GSMF). While the simulated GSMF is in good agreement with the observed one, some inconsistencies persist. In particular, we find the simulated GSMF to be slightly overpredicting the number density of low-mass galaxies at $z\gtrsim 2$.}


\maketitle

\section{Introduction} \label{sec:intro}

The increasing amount of multiwavelength data, obtained as part of wide-field and deep surveys, enables a progressively detailed examination of large samples of galaxies. Substantial advancements in determining important physical properties of galaxies (e.g., photometric redshifts, stellar masses, dust attenuation curves) have enabled the identification (in the rest-frame color-color space) of two separate types of galaxy populations: star-forming ones and quiescent ones (SFGs and QGs; e.g., \citealt{Labbe_2005, Wuyts_2007, Ilbert_2010}). The main difference between these two distinct populations of galaxies is their star formation activity; it has been shown that the star formation rate (SFR) and stellar mass ($M_\ast$) of the SFGs follow a tight correlation out to the highest redshifts probed -- the so-called star-forming main sequence (MS; e.g., \citealt{Speagle_2014, Tomczak_2016, Daddi_2022, Popesso_2023}). Most commonly, the SFR-$M_\ast$ relation can be parameterized with a simple power law:

\begin{equation}\label{eq:ms}
{\rm log(SFR/\Msun\,{\rm yr}^{-1})}=\alpha\, {\rm log}(M_\ast/\Msun)+\beta,
\end{equation}

\noindent where $\alpha$ and $\beta$ represent the slope and normalization of the MS, respectively. While the normalization has been shown to increase in redshift, with galaxies exhibiting higher SFRs at earlier epochs, the slope of the MS has been found to be close to unity, mostly independent of redshift, with values ranging between $\sim 0.8$ and 1.3 (e.g., \citealt{Lee_2015, Tomczak_2016, Popesso_2023}). More recently, a number of studies have found that the star-forming MS flattens at higher masses (e.g., \citealt{Lee_2015, Schreiber_2015, Leslie_2020, Leja_2022, Popesso_2023}), reaching an asymptotic value of the SFR. 

The exact shape and evolution of the MS is still widely discussed in the literature, with many studies presenting somewhat different perspectives. The sources of the discrepancies are many, including the assumed initial mass function (IMF), stellar population synthesis modeling, star formation histories (SFHs), and dust extinction curves. Most importantly, different techniques used for the estimation of the total SFRs can affect the results significantly. Since part of the UV emission coming from young stars gets absorbed by the interstellar dust, it is crucial to be able to properly quantify this effect. Measuring the dust attenuation from the rest-frame UV/optical data alone produces very uncertain results, because of the degeneracy between the age (different SFHs) and reddening (different metallicities, extinction curves). The advent of the {\it Herschel} Space Observatory \citep{Pilbratt_2010} allowed for the direct detection of the energy being re-emitted by dust in the far-infrared (FIR), and therefore a more reliable determination of the total (UV+IR) SFRs. However, since {\it Herschel} FIR data suffers from a relatively low resolution, only the most IR-luminous sources can be detected at high redshifts, and hence samples of SFGs studied with {\it Herschel} tend to be incomplete, often biasing the results toward large values of the SFRs. \citet{Speagle_2014} and, more recently, \citet{Popesso_2023} compiled various studies in the literature in order to convert them to the same absolute calibrations in an attempt to solve some of the issues listed above. While they carefully address all the inconsistencies caused by different modeling assumptions, the uncertainties associated with the various total SFR indicators, as well as the completeness of samples compiled, remain a concern.

In this work, we make use of the sample of $\sim 100$k $K$-band-selected galaxies from the ground-based UV+optical+near-IR $\sim 2$\,deg$^2$ imaging of the UKIDSS Ultra Deep Survey (UDS) and COSMOS fields \citep{McLeod_2021}. Galaxies are stacked in bins of redshift and stellar mass in the FIR {\it Herschel} and the {\it James Clerk Maxwell} Telescope (JCMT) SCUBA-2 \citep{Holland_2013} maps, in order to determine the shape and redshift evolution of the star-forming MS out to $z\lesssim 5.7$. The stacking approach, used by several authors before (e.g., \citealt{Tomczak_2016, Leslie_2020, Merida_2023}), has several advantages. The sample of galaxies stacked is mass-complete, assuring that the resulting values of the IR luminosity (and SFR) are not biased due to selection effects. In addition, thanks to the inclusion of the SCUBA-2 850\,${\rm \mu m}$ data, our stacked photometry maps the Rayleigh–Jeans portion of the dust emission spectrum, allowing for the precise determination of the  dust temperatures, and thus the SFRs. 

The paper is structured as follows. In Sect. \ref{sec:data}, the data used in this work is described. We outline the stacking procedure and the star-forming sample selection process in Sect. \ref{sec:stack}. Section\,\ref{sec:ms} explains the methodology used in the determination of the star-forming MS for our sample. We discuss the results of this work in Sect. \ref{sec:disc}. In particular, the shape and evolution of the star-forming MS and its relation to the so-called Schmidt-Kennicutt relation (SK relation; \citealt{Schmidt_1959, Kennicutt_1998}) is explained in Sect. \ref{sec:ev}, followed by a discussion of the variations in the peak dust temperature with IR luminosity and redshift in Sect. \ref{sec:dust} and a comparison to other works in Sect. \ref{sec:comp}. In Sect. \ref{sec:gsmf}, we attempt to interpret some of the discrepancies between the different shapes of the MS found in the literature, by “growing” the observed galaxy stellar mass function (GSMF) in time. Finally, we summarize our conclusions in Sect. \ref{sec:sum}. Throughout the paper, we use the \citet{Chabrier_2003} stellar IMF with an assumed flat cosmology of $\Omega_{\rm m} = 0.3$, $\Omega_\Lambda = 0.7$, and ${\rm H}_0 = 70$\,km\,s$^{-1}$\,Mpc$^{-1}$.

\section{Data} \label{sec:data}

\subsection{Optical/near-IR catalogs} \label{sec:opt}

We started with the optical/near-IR catalogs of the UKIDSS Ultra Deep Survey (UDS) and COSMOS fields. The details of the data used, galaxies selection process and determination of photometric redshifts and stellar masses are presented in \citet{McLeod_2021}, with the shortened description given below.

\subsubsection{Imaging}

For the UKIDSS UDS field \citep{Lawrence_2007}, we included UDS DR11 imaging (Almaini et al., in preparation) in the near-IR JHK bands, with Y-band imaging from the VISTA VIDEO DR4 release over XMM-LSS \citep{Jarvis_2013}. We complemented this with UV imaging in $u^\ast$-band from CFHT MegaCam, and optical imaging in BVRiz$^{\prime}$ from Subaru Suprime-Cam \citep{Furusawa_2008}. We also employed imaging in the narrowband NB921 \citep{Koyama2011} and the refurbished Suprime-Cam $z$-band ($z'_{\rm new}$; \citealt{Furusawa_2016}). To extend our wavelength coverage into the mid-infrared, we utilized \textit{Spitzer} IRAC imaging in 3.6\,$\mu$m and 4.5\,$\mu$m, combining the SEDS \citep{Ashby_2013}, S-CANDELS \citep{Ashby_2015}, and SPLASH (PI Capak; see \citealt{mehta2018}) programs. The effective area of the overlapping UDS imaging was 0.69 deg$^2$, after masking for bright stars.

For the COSMOS field, we combined near-IR $YJHK_{\rm s}$ imaging from UltraVISTA DR4 \citep{McCracken_2012} with UV/optical $u^\ast griz$ imaging from the CFHTLS-D2 T0007 data release \citep{hudelot2012}. We confine our study to the 1 deg$^2$ overlapping area covered by these data sets. After masking, this reduced to an effective area of 0.86\,deg$^2$. As with the UDS, we also included Subaru Suprime-Cam z$^{\prime}_{new}$ imaging. There is a wealth of \textit{Spitzer} IRAC imaging available in 3.6\,$\mu$m and 4.5\,$\mu$m over the COSMOS field. The programs we combined include S-COSMOS (Sanders et al. 2007), SPLASH (PI Capak), SEDS (Ashby et al. 2013), S-CANDELS (Ashby et al. 2015) and SMUVS (Ashby et al. 2018).

\begin{figure*}
\centering
   \includegraphics[width=17cm]{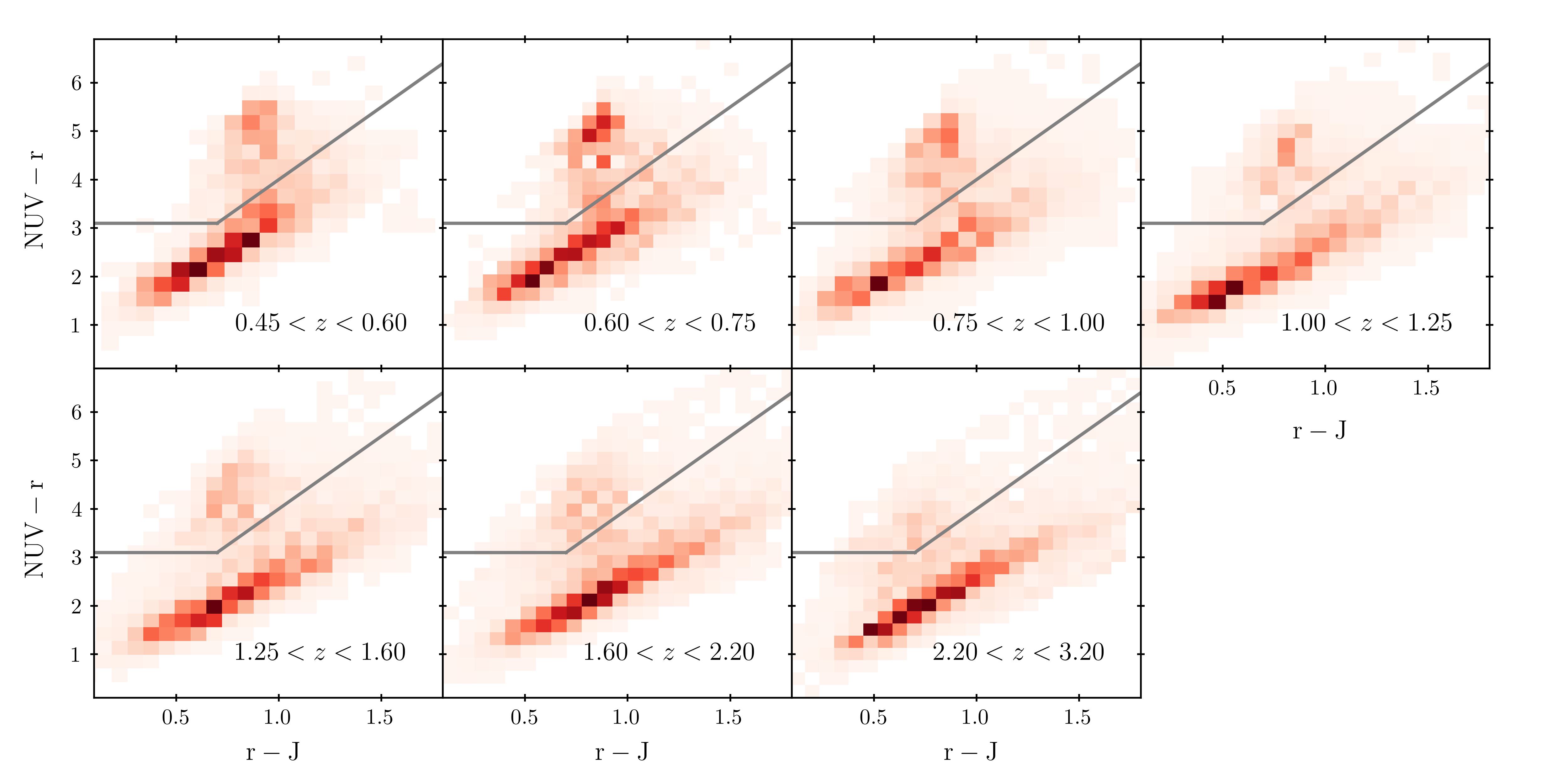}
     \caption{Results of the $NUVrJ$ color-color selection results, defined in Eq. \ref{eq:nuvrj} shown for redshift bins adopted in this work. The $NUVrJ$ colors \citep{Ilbert_2013} are used in order to distinguish between sources reddened due to the lack of star formation (quiescent galaxies -- upper left section of the plots) from those where red colors are caused by dust extinction (SFGs -- main locus in the plots).}
     \label{fig:nuvrj}
\end{figure*}

\subsubsection{Photometric redshifts and stellar masses}

As is explained in detail in \citet{McLeod_2021}, the parent multiwavelength catalogs were constructed by running {\sc Source-Extractor} \citep{Bertin_1996} in dual-image mode, with the K-band as the detection image. Photometry was measured in 2-arcsec diameter apertures on the PSF-homogenized UV to near-IR images. For the lower resolution {\it Spitzer} IRAC 3.6 and 4.5\,${\rm \mu m}$ data, the photometry was measured using the software {\sc tphot} \citep{Merlin_2015}.

For each object in the catalog, the photometric redshift we assigned was the median of six SED fitting runs, employing three different software codes. With {\sc LePhare}, we included runs with the \citet{Bruzual_2003}, \textsc{Pegase} \citep{Fioc_1999} and COSMOS \citep{Ilbert_2009} template libraries. We also included  two {\sc eazy} \citep{Brammer_2008} runs, using the PCA and \textsc{Pegase} templates, and one further run with the BPZ code \citep{Benitez_2000}, using the \citet{Coleman_1980} templates. The resulting values of $\sigma_z$, defined as $1.483 \times {\rm MAD}(dz)$, where MAD is the median absolute deviation and $dz = (z_{\rm phot} - z_{\rm spec})/(1 + z_{\rm spec})$, for COSMOS and UDS fields were found to be equal to 0.019 and 0.022, respectively \citep{McLeod_2021}.

Stellar masses were measured for each object by fixing the photometric redshift to $z_{\rm med}$, and re-fitting the SED using {\sc LePhare} with the \citet{Bruzual_2003} library. This template set included a \citet{Chabrier_2003} IMF, a \citet{Calzetti_2000} dust attenuation law and IGM absorption as in \citet{Madau_1995}. The values were found to be in excellent agreement with those derived by the CANDELS team (see Table 2 of \citealt{McLeod_2021}), with a tight 1:1 relation and a typical scatter of $\pm 0.05$ dex. The resulting catalogs were then cleaned of stars, AGN, artifacts and objects with contaminated photometry. Finally, a further $\chi^{2}<40$ goodness-of-fit cut was implemented on the catalogs.

\subsection{Far-infrared} \label{sec:fir}

For the extraction of the stacked FIR flux densities in the COSMOS and UDS fields we used the {\it Herschel} \citep{Pilbratt_2010} Multi-tiered Extragalactic Survey (HerMES; \citealt{Oliver_2012}) and the Photodetector Array Camera and Spectrometer (PACS; \citealt{Poglitsch_2010}) Evolutionary Probe (PEP; \citealt{Lutz_2011}) data obtained with the Spectral and Photometric Imaging Receiver (SPIRE; \citealt{Griffin_2010}) and PACS instruments. We utilized Herschel maps at 100, 160, 250, 350 and 500\,${\rm \mu m}$ with beam sizes of 7.39, 11.29, 18.2, 24.9, and 36.3\,arcsec, and 5$\sigma$ sensitivities of 7.7, 14.7, 24.0, 27.5, and 30.5\,mJy, respectively.

In order to constrain the Rayleigh–Jeans tail of the dust emission curve, we have also included the data collected as a part of the SCUBA-2 Cosmology Legacy Survey (S2CLS; \citealt{Geach_2017}), with the UKIDSS-UDS  and COSMOS 850\,${\rm \mu m}$ imaging covering $\simeq0.9$\,deg$^2$ with a 1$\sigma$ noise of 0.9\,mJy and  $\simeq1.3$\,deg$^2$ with the 1$\sigma$ noise of 1.6\,mJy, respectively.

\section{Stacking} \label{sec:stack}

\subsection{Sample selection} \label{sec:ss}

For the purpose of successful stacking (i.e., to maximize the number of sources in each stack), we designed each redshift bin to encompass $\sim 1$ billion years of Universe evolution, ranging from 1 to 9 billion years after the Big Bang (with the corresponding redshift bins of [0.45, 0.6, 0.75, 1.0, 1.25, 1.6, 2.2, 3.2, 5.7]; see Table\,\ref{tab:sfr}). We also made sure that the sample we stacked is mass-complete. The procedure for estimating the mass completeness in the UDS and COSMOS fields is explained in \citet{McLeod_2021}. In brief, the completeness was estimated using simulations, in which $K$-band imaging was produced by injecting a set of artificial galaxies with a wide range of physical properties. Adopting the original setup, the corresponding photometry catalogs were then constructed using {\sc sextractor}. In order to determine the 90\% mass-completeness limit, the relation from \citet{Pozzetti2010} was used to derive the limiting stellar mass for each object in our sample. The 90\% mass completeness limit, below which 90\% of the limiting stellar masses lie, could then be derived as a function of redshift. The resulting redshift and stellar mass bins adopted in this work are listed in Table\,\ref{tab:sfr}.

In order to derive the SFR-$M_\ast$ relation for the star-forming sample, we first needed to separate it from the quiescent population. Here, we adopted the selection based on the $NUVrJ$ colors \citep{Ilbert_2013}:   

\begin{align}\label{eq:nuvrj}
\begin{split}
(NUV-r)> &\, 3\times(r-J) + 1;\\
(NUV-r)> &\, 3.1,
\end{split}
\end{align}

\noindent where $NUV$, $r$ and $J$ rest-frame magnitudes were generated from the best-fit SEDs. The advantage of this method is that it is able to separate sources for which reddening is caused by aging from those for which it is produced by dust extinction, thereby decreasing the chance of contaminating the dust-enshrouded star-forming sample with quiescent galaxies. Because at high redshifts the rest-frame $J$-band can no longer be traced by the photometry available in this work, the $NUVrJ$ color-color selection was performed out to $z\sim 4$, as is depicted in Fig. \ref{fig:nuvrj}.

To avoid overestimating the SFRs at the high-mass end of the MS, we further detected and removed the starburst galaxy population from our stacking sample. To identify starbursts, we used the ALMA-detected sources in the COSMOS \citep{Liu_2019} and UDS \citep{Stach_2019} fields, with associated stellar masses and SFRs, as has been found by \citet{Liu_2019} and \citet{Dudzeviciute_2020}, respectively. Following \citet{Elbaz_2018}, starbursts are defined as galaxies with ${\rm SFR/SFR_{MS}}>3$, where ${\rm SFR_{MS}}$ is the corresponding MS value, given the redshift and stellar mass of each ALMA-detected source, with the functional form of the star-forming MS of \citet{Popesso_2023}.

\subsection{Far-infrared fluxes} \label{sec:sfir}

Since current FIR/submillimeter surveys struggle to reach the (sub-)millijansky depths necessary for the detection of the dust emission in typical SFGs, we have resorted here to the well-known method of stacking (e.g., \citealt{Whitaker_2014, Schreiber_2015, Tomczak_2016, Koprowski_2018}). In order to determine the FIR fluxes at a given band, we cut a square stamp around the position of each source, with the size of approximately 3 $\times$ the full width at half maximum (FWHM) of the corresponding beam. The average FIR fluxes at each pixel were then evaluated following the inverse-variance weighting:

\begin{equation}\label{eq:stack}
S_\nu=\frac{\sum_i S_{\nu,i}/\sigma_i^2}{\sum_i 1/\sigma_i^2},
\end{equation} 

\noindent where $S_{\nu,i}$ is the flux density of the $i$th source and $\sigma_i$ is the 1$\sigma$ instrumental noise, extracted at each pixel in a given source's stamp. The error on the mean stacked fluxes for each pixel was calculated according to

\begin{equation}\label{eq:dstack}
\sigma_{\rm stack}^2(S_\nu)=\frac{1}{\sum_i 1/\sigma_i^2},
\end{equation}

\noindent with the example 500\,${\rm \mu m}$ stamps for each redshift and stellar mass bin given in Sect. \ref{sec:datav}.

\begin{figure}
\centering
   \includegraphics[width=7.5cm]{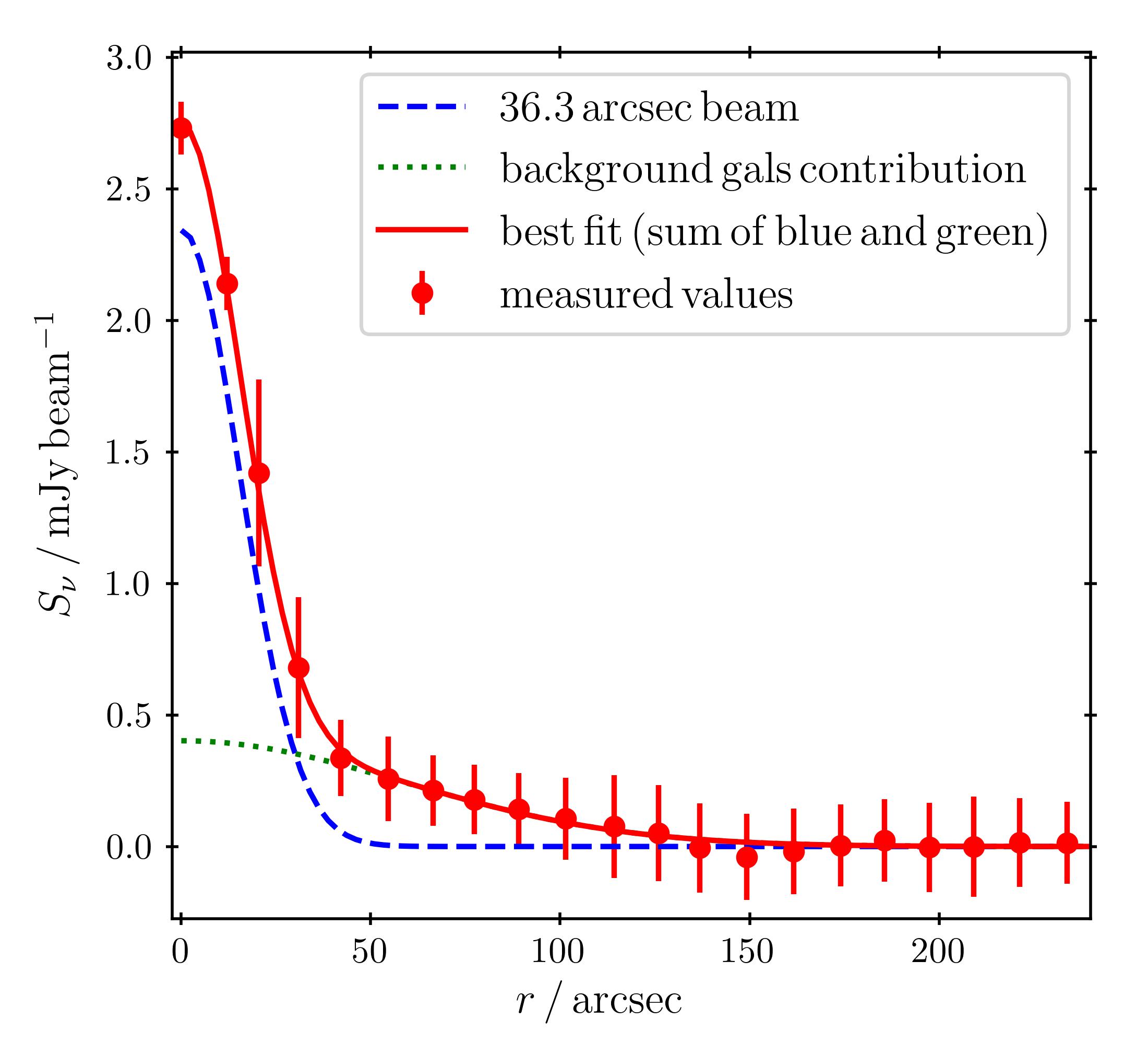}
     \caption{$1.25<z<1.75$, $10.25<{\rm log}(M_\ast/{\rm M_\odot})<10.5$ light profile derived at 500\,${\rm \mu m}$. The red points represent mean values of the flux densities derived from stacked images as a function of distance from the image center. The dashed blue, dotted green, and solid red lines represent the corresponding beam, the contribution from background sources, and the sum of the two, respectively. For {\it Herschel} SPIRE bands, the contribution from background sources is significant and must therefore be subtracted from the overall light profile, when deriving stacked flux densities (see Sect. \ref{sec:sfir}).}
     \label{fig:cluster}
\end{figure}

The stacked average FIR flux densities at each bin were then extracted from the nonlinear least squares best fits to the light profiles derived from corresponding stamps. The reason for this is that, at the {\it Herschel} 250, 350, and 500\,${\rm \mu m}$ SPIRE bands, due to the relatively large beam sizes, the clustering of objects associated with the stacked sources is seen to significantly contribute to the mean flux. Figure\,\ref{fig:cluster} shows the light profile and the best fits for the sample of {\it Herschel} SPIRE 500\,${\rm \mu m}$ sources at $1.25<z<1.75$ with stellar masses between $10.25<{\rm log}(M_\ast/{\rm M_\odot})<10.5$. As can be seen, the stacked flux (red line) consists of the average flux from the stacked sample (dashed blue line) and the background flux from the associated sources (dotted green line), both of which can be modeled with a Gaussian function. In cases when a double Gaussian (${\rm beam + background}$) gives better fits than a single Gaussian (beam), the background is subtracted and the resulting average flux density of the stacked sample is extracted from a beam component of the best fit. 

The errors on the stacked FIR fluxes, which are related to the redshifts, stellar masses, and the $NUVrJ$ quiescent galaxy selection procedure uncertainties, were estimated using the bootstrapping method. In each step, a mock catalog was constructed by drawing, at random and with replacement, a sample of sources from the original mass-complete dataset. The stacking procedure, described above, was then applied and the average fluxes were extracted. The process was repeated 1000 times and the uncertainties on the FIR fluxes, $\sigma_{\rm bootstrap}$, taken to be the standard deviation of the resulting simulated values. In addition, the contribution to the final stacked FIR fluxes errors from the uncertainties related to the functional fits to the light profiles, $\sigma_{\rm fit}$, described above, as well as the errors related to the individual observed FIR fluxes from Eq. \ref{eq:dstack} were included, with the final errors on the stacked values equal to

\begin{equation}\label{eq:dstackfin}
\sigma(S_\nu)=\sqrt{\sigma_{\rm bootstrap}^2+\sigma_{\rm fit}^2+\sigma_{\rm stack}^2}.
\end{equation}

% \begin{figure}
% \centering
%   \resizebox{0.7*\hsize}{!}{\includegraphics{figures/cluster.jpeg}}
%   \caption{The $1.25<z<1.75$, $10.25<{\rm log}(M_\ast/{\rm M_\odot})<10.5$ light profile derived at 500\,${\rm \mu m}$. The red points represent mean values of the flux densities derived from stacked images as a function of distance from the image center. The blue dashed, green dotted and red solid lines represent the corresponding beam, the contribution from background sources and the sum of the two, respectively. For {\it Herschel} SPIRE bands, the contribution from background sources is significant and must, therefore, be subtracted from the overall light profile, when deriving stacked flux densities (see Section\,\ref{sec:sfir}).}
%   \label{fig:cluster}
% \end{figure}

\section{The star formation rate--stellar mass relations} \label{sec:ms}

\subsection{Star formation rate measurements} \label{sec:sfrm}

A total SFR consists of the UV and IR components, where the rest-frame UV light is sensitive to the young massive stars and is therefore a good tracer of recent, unobscured star formation, while the IR emission accounts for the stellar emission that has been absorbed by the interstellar dust.  Here, we followed the recent calibration of \citet{Kennicutt_2012}, where

\begin{equation}\label{eq:sfrtot}
{\rm SFR_{tot}}=1.71\times 10^{-10}(L_{\rm FUV}+0.46\times L_{\rm IR}).
\end{equation}

\noindent The far-UV (FUV) part of the SFR was derived from the FUV luminosity measured at 150\,nm, where $L_{\rm FUV}=\nu L_\nu$, with FUV SFRs calculated for each galaxy and then the mean value evaluated at each redshift or stellar mass bin.

\begin{figure*}
\centering
   \includegraphics[width=17cm]{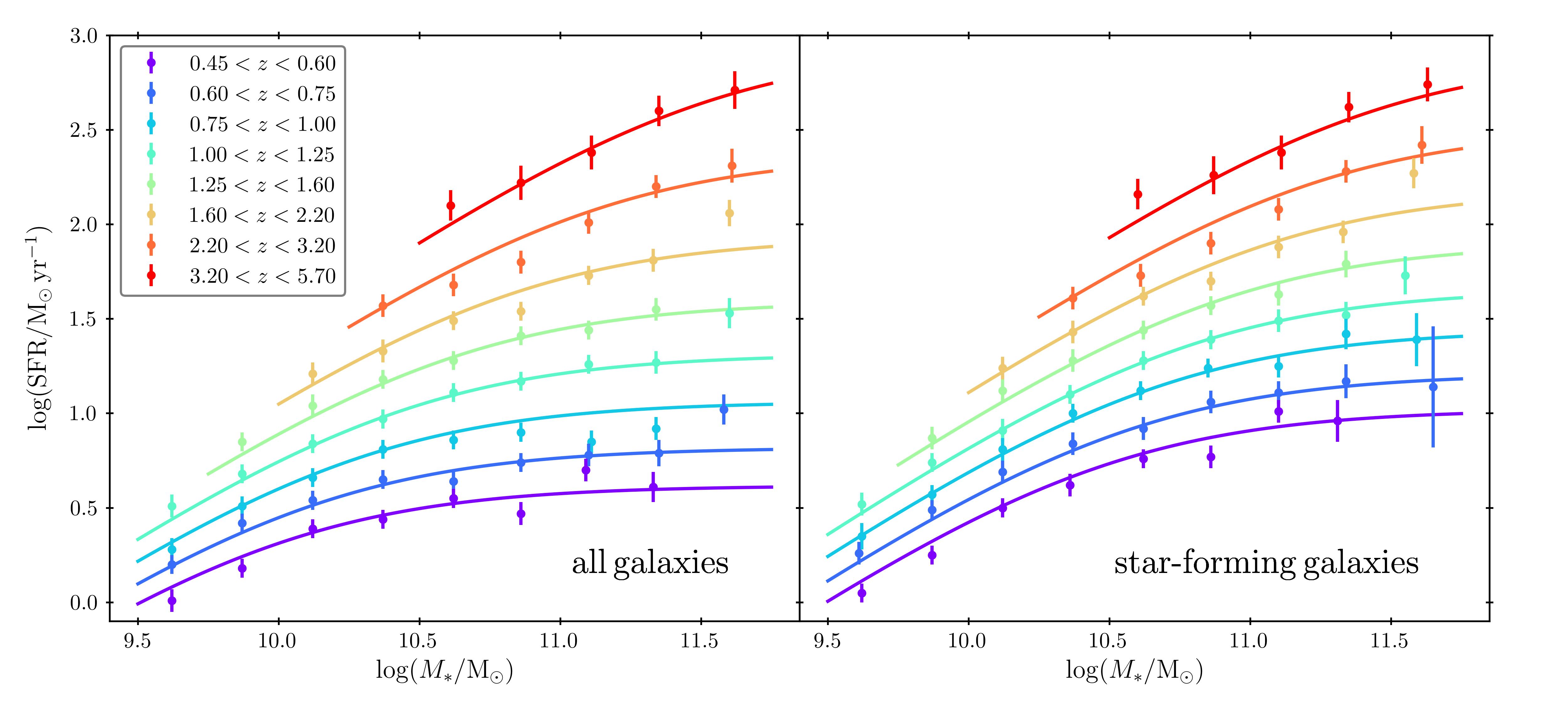}
     \caption{SFR-$M_\ast$ relations derived in Sect. \ref{sec:ms} for the whole (star-forming and quiescent -- left panel) and the star-forming (MS -- right panel) galaxy samples. The best-fit lines, color-coded with redshift, were derived according to Eq. \ref{eq:sfr2}.}
     \label{fig:sfr}
\end{figure*}

In order to find the IR luminosity, $L_{\rm IR}$, the dust emission curve of \citet{Casey_2012} was fit to the available FIR photometry using the nonlinear least squares analysis. The adopted dust emission curve, described in Table\,1 of \citet{Casey_2012}, can be approximated as a classic single-temperature gray-body function at FIR wavelengths, with the addition of the power law fit at mid-IR wavelengths. The data was initially fit with both the single gray-body and the power law gray-body curves and it was found that the latter gives better fits to the data at mid-IR bands. The power law gray-body curve has four free parameters -- the mid-IR power law slope, $\alpha$, the emissivity index, $\beta$, the effective dust temperature, $T^{\rm eff}_{\rm d}$, and the overall normalization, $N$. Since, at $z\gtrsim 2$, the available photometry redward of the dust emission peak consists only of the 850 and 500\,${\rm \mu m}$, with the 500\,${\rm \mu m}$ being affected by blending (see Fig. \ref{fig:cluster}), we decided to fix $\beta$ at 1.96. Similarly, due to the lack of photometry at a rest-frame wavelength of $\lambda \lesssim 80\,{\rm \mu m}$, we also fixed $\alpha$ at 2.3. Values of $\alpha$ and $\beta$ were fixed according to the recommendations of \citet{Drew_2022}. The curve was then integrated between 8 and 1000\,${\rm \mu m}$ and the errors were estimated from the corresponding covariance matrix. The stacked photometry and the best fits for the whole (star-forming and quiescent) and star-forming subsamples are given in Sect. \ref{sec:datav}. The resulting best-fit values of the total SFR at each redshift or stellar mass bin for the whole and star-forming subsamples are listed in the top and bottom panels of Table\,\ref{tab:sfr}, as well as being depicted as color points with error bars in the left and right panels of Fig. \ref{fig:sfr}, respectively.

\subsection{Fitting the star formation rate--stellar mass function} \label{sec:fit}

Since the SFR-$M_\ast$ relation was shown to exhibit a turnover at high masses (e.g.,~\citealt{Lee_2015, Popesso_2023}), we used the functional form of the SFR-$M_\ast$ relation from \citet{Lee_2015}:

\begin{equation}\label{eq:sfr}
{\rm SFR} = {\rm SFR_{max}}/(1+(M_\ast/M_0)^{-\gamma}),
\end{equation}

\noindent where the SFR transitions from a power law with slope $\gamma$ at masses below $M_0$ to an asymptotic value, ${\rm SFR_{max}}$, at $M_\ast\gg M_0$. The slope, $\gamma$, has been shown to vary with redshift, with values between $0.8\lesssim \gamma \lesssim 1.3$ (e.g., \citealt{Lee_2015, Tomczak_2016}). Because of the low dynamic range in $M_\ast$, particularly at high redshift, we decided to fix $\gamma$. With the best-fit value of $\gamma=1.0$, adopted from \citet{Popesso_2023}, Eq. \ref{eq:sfr} becomes

\begin{equation}\label{eq:sfr2}
{\rm SFR} = {\rm SFR_{max}}/(1+(M_0/M_\ast)).
\end{equation}

\noindent We find that the most accurate fits to the SFR-$M_\ast$ relation are produced when the logarithm of the free parameters of Eq. \ref{eq:sfr2} are assumed to follow an exponential evolution with redshift, where

\begin{equation}\label{eq:pars}
\begin{split}
{\rm log(SFR_{max}/M_\odot yr^{-1})} & = a_1+a_2\times e^{-a_3z} \\
{\rm log}(M_0/{\rm M_\odot}) & = b_1+b_2\times e^{-b_3z}.
\end{split}
\end{equation}

\begin{table}
\caption{Best-fit values to parameters from Eq. \ref{eq:pars} for the star-forming galaxies and for all galaxies (both star-forming and quiescent).}\label{tab:pars}
\centering
\begin{tabular}{ccc}
\hline\hline
Parameter & All galaxies & SF galaxies \\
\hline
a1 & $\phantom{-1}3.28\pm 0.32$ & $\phantom{-1}2.97\pm 0.20$ \\
a2 & \phantom{1}$-3.36\pm 0.26$ & \phantom{1}$-2.62\pm 0.14$ \\
a3 & $\phantom{-1}0.46\pm 0.08$ & $\phantom{-1}0.59\pm 0.10$ \\
\hline
b1 & $\phantom{-}11.73\pm 0.73$ & $\phantom{-}11.38\pm 0.39$ \\
b2 & \phantom{1}$-2.06\pm 0.62$ & \phantom{1}$-1.14\pm 0.30$ \\
b3 & $\phantom{-1}0.37\pm 0.23$ & $\phantom{-1}0.50\pm 0.36$ \\
\end{tabular}
% \tablecomments{This table ``hides'' the third column in the \latex\ when compiled.
% The Distance is also centered on the decimals.  Note that when using decimal
% alignment you need to include the {\tt\string\decimals} command before
% {\tt\string\startdata} and all of the values in that column have to have a
% space before the next ampersand.}
\end{table}

\noindent The functional form of Eq. \ref{eq:sfr2} was fit to the stacked data (Table\,\ref{tab:sfr}) using the nonlinear least square fitting, where, initially, the fits were performed for the whole sample, including both star-forming and quiescent galaxies. The best-fit curves can be seen in the left panel of Fig. \ref{fig:sfr} and the best-fit parameters from Eq. \ref{eq:pars} for the whole sample are listed in Table\,\ref{tab:pars}.

As can be seen in Fig. \ref{fig:sfr}, the SFR-$M_\ast$ relation shows a clear turnover at high stellar masses, represented in Eq. \ref{eq:sfr2} by the parameter $M_0$. However, since the whole sample consists of both the star-forming and quiescent galaxies, the apparent turnover may simply be a consequence of the increasing population of quiescent galaxies at low redshifts. We therefore repeated the procedure for the SFGs only. The corresponding best SFR-$M_\ast$ fits are shown in the right panel of Fig. \ref{fig:sfr}, with the best-fit parameters listed in Table\,\ref{tab:pars}. As can be seen, the turnover is still visible at low redshifts, albeit slightly less pronounced.

\section{Discussion}\label{sec:disc}

\begin{figure}
\centering
   \includegraphics[width=7.5cm]{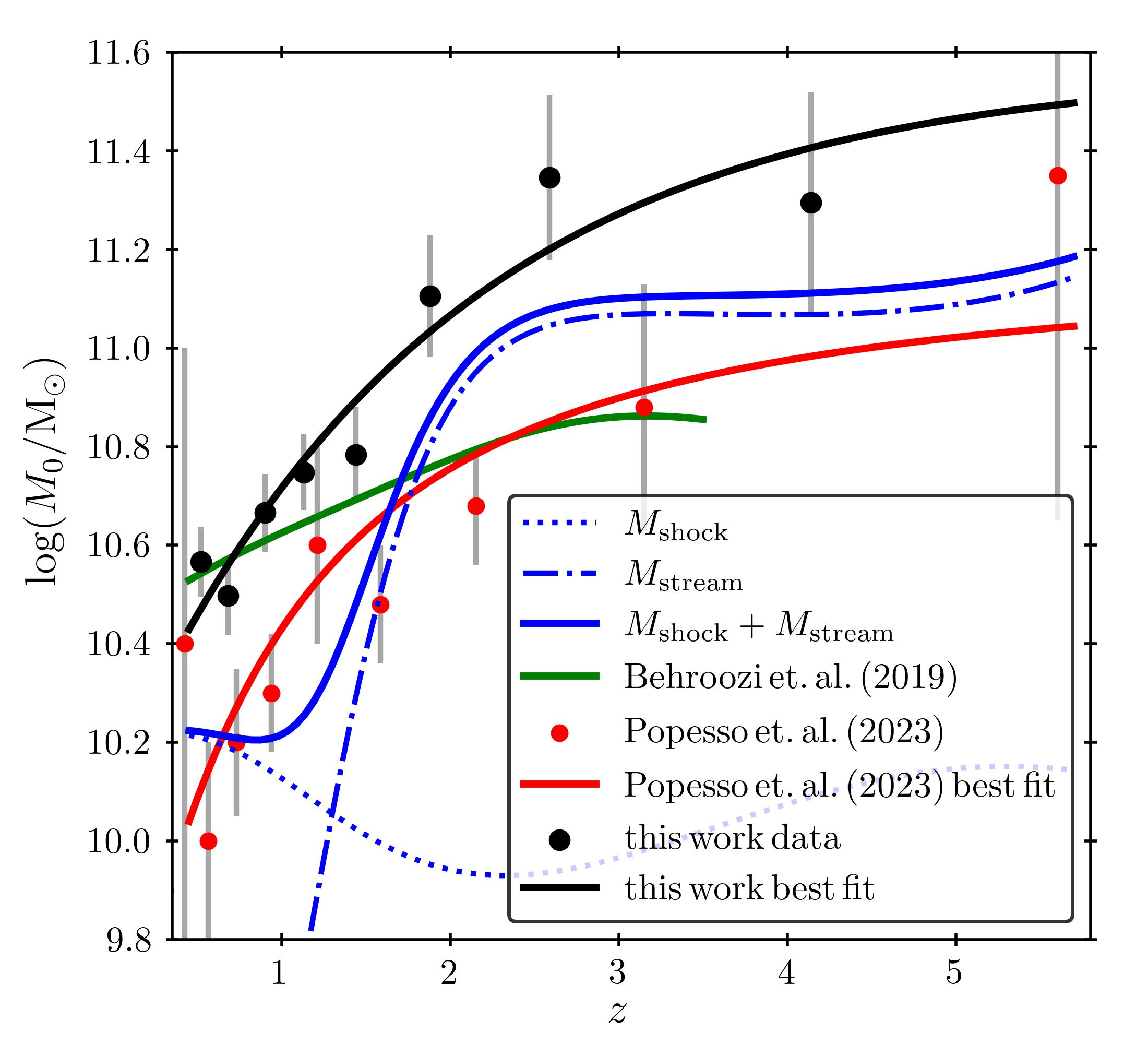}
     \caption{$M_{\rm shock}$ and $M_{\rm stream}$ boundaries, above which the star formation is theoretically expected to be quenched, as is described in Sect. \ref{sec:ev}, shown in blue, where halo masses were converted to stellar masses using the prescription of \citet{Behroozi_2013}. The empirical curve of \citet{Behroozi_2013} is also shown with a solid green line. For comparison, the results of this work are plotted in black. Red data represents the $M_0$ redshift evolution found in \citet{Popesso_2023}.}
     \label{fig:m0}
\end{figure}

\subsection{Evolution of the star-forming main sequence}\label{sec:ev}

As has been discussed in previous studies (e.g., \citealt{Santini_2014, Wang_2022}), the shape of the star-forming MS is controlled by the relationship between the molecular gas mass and SFR surface densities -- the SK relation. After stacking a mass-complete sample of $0.4\lesssim z\lesssim 3.6$, $M_\ast>10^{10}\,{\rm M_\odot}$ MS galaxies, \citet{Wang_2022} found that all the MS galaxies lie on the stellar mass and redshift independent MS-only SK relation, with the slope of $\sim 1.13$. For comparison, at high $\Sigma M_{\rm gas}$, starburst galaxies exhibit two to three times higher star-forming efficiencies, producing the SK relation with slightly higher normalization (e.g., \citealt{Daddi_2010, Kennicutt_2021}). In addition, \citet{Wang_2022} showed that the FIR sizes of all the MS galaxies do not vary with stellar mass or redshift, with the mean half-light radius of $\sim 2.2$\,kpc, implying the so-called integrated SK law, where ${\rm SFR}\propto M_{\rm gas}^{1.13}$. The star-forming MS stellar mass and redshift evolution is therefore driven by the shape of the integrated SK relation and galaxy molecular gas content ($f_{\rm gas}$). The molecular gas fraction, $f_{\rm gas}\equiv M_{\rm gas}/M_\ast$, on the other hand, is driven by many processes, including the star formation efficiency (${\rm SFE\equiv SFR}/M_{\rm gas}$), gas cooling, feedback, and the merger rate. As has been shown by \citet{Santini_2014}, the redshift and stellar mass evolution of the molecular gas content in SFGs, together with the redshift-independent SK relation, give rise to the fundamental $f_{\rm gas}$-$M_\ast$-SFR three-dimensional surface. \citet{Santini_2014} stacked  GOODS-S, GOODS-N and the COSMOS SFGs out to $z=2.5$ in {\it Herschel} bands to show that the molecular gas fraction at given stellar mass and SFR does not depend on redshift. The projection of the $f_{\rm gas}$-$M_\ast$-SFR surface onto the $M_\ast$-SFR plane, given $f_{\rm gas}$ variations with stellar mass and redshift, produces the evolution of the star-forming MS.

From the best-fit function describing the MS evolution below the bending mass $M_0$ (Eqs.\,\ref{eq:sfr} and \ref{eq:pars}) at any given redshift, the SFR dependence on the stellar mass is close to linear ($\gamma\sim 1$), consistent with the constancy of the low-mass power law slope of the stellar mass function of $\alpha\sim 1.5$ (e.g., \citealt{Tomczak_2014}, although see Sect. \ref{sec:gsmf}). Since, from the MS-only integrated SK relation, ${\rm SFR}\propto M_{\rm gas}^{1.13}$, a linear dependence of the SFR on stellar mass implies that at a given redshift and below $M_0$, the gas fraction should decrease with stellar mass, as was indeed found in previous works (e.g., \citealt{Schreiber_2016, Wang_2022}). Similarly, since ${\rm SFR}\propto M_\ast$ below $M_0$ and the gas fraction decreases with stellar mass, the star-forming efficiency ($\equiv {\rm SFR}/M_{\rm gas}$) will increase with $M_\ast$. On the other hand, for a given stellar mass, increasing the SFR with redshift (evolution of the MS normalization) should also increase the molecular gas mass (MS-only SK relation). For a given stellar mass, the gas fraction must therefore increase with redshift, which is consistent with previous findings (e.g., \citealt{Geach_2011, Wang_2022}).

At masses $>M_0$, the MS flattens. As has been suggested (e.g., \citealt{Daddi_2022, Popesso_2023}), the SFR-$M_\ast$ turnover is most likely caused by the phasing out of cold accretion. Based on theoretical scenarios, the cold gas can accrete freely for dark matter halos with $M_{\rm DM}<M_{\rm shock}$, where $M_{\rm shock}$ ($\simeq 10^{11.8}\,{\rm M_\odot}$ independent of redshift) marks the mass above which shocks can efficiently heat incoming baryons. Above $M_{\rm shock}$, the cold accretion can still occur at higher redshifts through cold, collimated streams for halos with $M_{\rm DM}<M_{\rm stream}$, with $M_{\rm stream}$ boundary evolving with redshift from $\sim 10^{12.5}\,{\rm M_\odot}$ at $z=2$ to $\sim 10^{13.5}\,{\rm M_\odot}$ at $z=3$. Converting from halo to stellar masses using the prescription of \citet{Behroozi_2013}, we plot the redshift evolution of the $M_{\rm shock}$-$M_{\rm stream}$ boundary in Fig. \ref{fig:m0}, with the $M_{\rm shock}$, $M_{\rm stream}$ and $M_{\rm shock}+M_{\rm stream}$ boundaries shown in dotted, dash-dotted, and solid blue lines, respectively. We also plot the results of this work in black, where it can be seen that the boundary is in good agreement with the redshift evolution of the SFR-$M_\ast$ turnover mass $M_0$ (increasing exponentially toward higher redshifts; Eq. \ref{eq:pars}). For comparison, the results of \citet{Behroozi_2013} and \citet{Popesso_2023} are also shown. This suggests that the flattening of the SFR-$M_\ast$ relation, at a given redshift, is most likely caused by the suppression of the cold-gas feeding of the galaxy, which effectively decreases its molecular gas fraction. At the same time, \citet{Daddi_2022} points out that this process is not equivalent to quenching, describing galaxies with very low SFRs, for which additional processes are required, the most likely candidates of which include mergers and AGNs. 

Flattening of the MS at $M_\ast>M_0$ gives a relatively small increase in SFR for a large increase in stellar mass. Provided that MS galaxies lie on the integrated SK relation, a small increase in SFR gives a small increase in molecular gas mass, causing the molecular gas fractions above $M_0$ to decrease significantly, consistent with the above scenario. However, \citet{Schreiber_2016} found that flattening of the star-forming MS at $z\sim 1$ is caused not by the decrease in the molecular gas fraction but rather by the significant drop in the star-forming efficiency, with only a small change in $f_{\rm gas}$. For galaxy stellar masses above the bending mass, the large increase in $M_\ast$ gives only a small increase in SFR. A small change in $f_{\rm gas}$ would imply that the corresponding molecular gas mass should also significantly increase. This is inconsistent with the finding of \citet{Wang_2022} that all the SFGs lie on the SK relation, where ${\rm SFR}\propto M_{\rm gas}^{1.13}$, since a small increase in the SFR above $M_0$ for MS galaxies should give similarly small rise in the molecular gas mass (and hence large decrease in $f_{\rm gas}$). A near-constant gas fraction above $M_0$ would therefore suggest that the MS galaxies at these large masses should lie below the SK relation. More work, with larger samples, is clearly required to resolve this inconsistency.

\begin{figure*}
\centering
   \includegraphics[width=17cm]{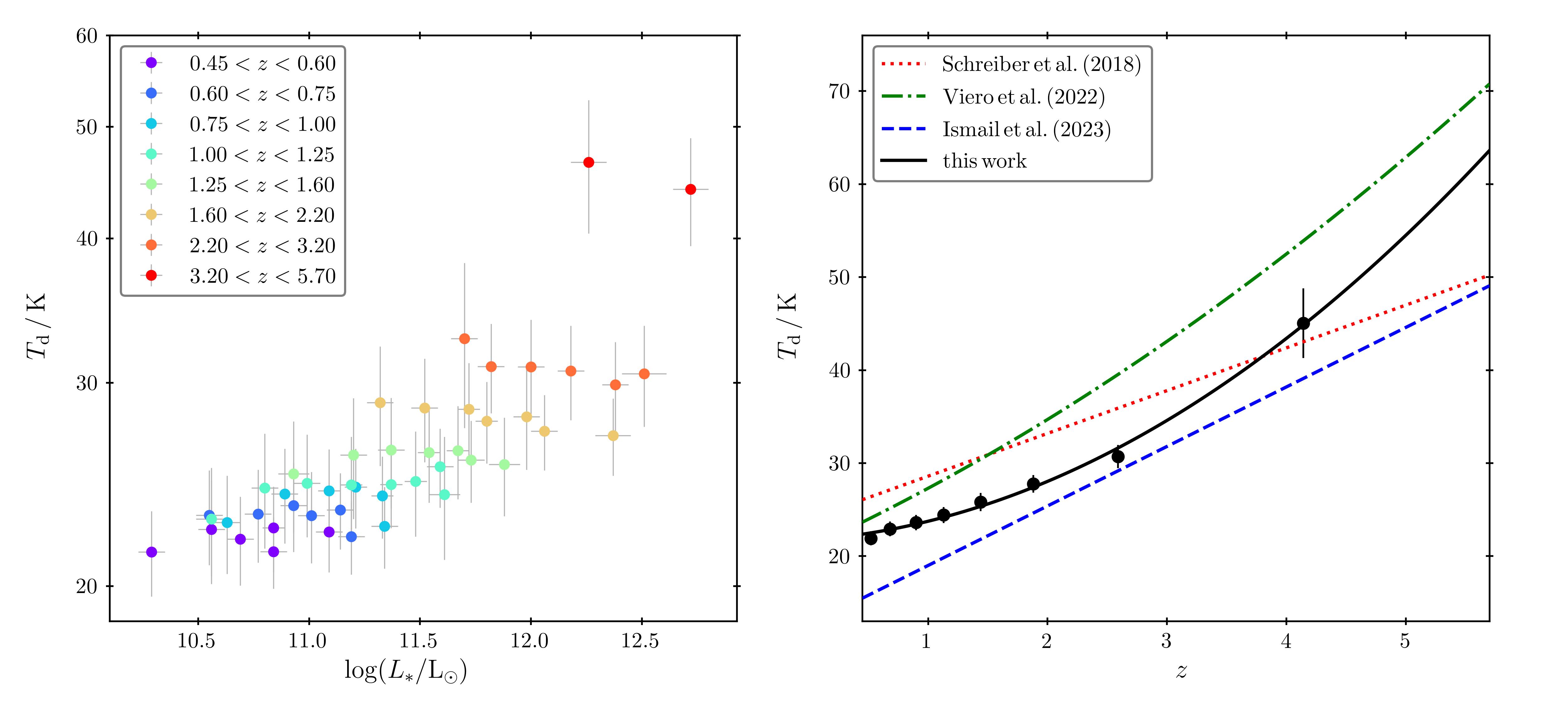}
     \caption{Evolution of peak dust temperature found in this work. {Left panel}: The peak dust temperatures, $T_{\rm d}$, as a function of the IR luminosity, $L_{\rm IR}$, color-coded with redshift. It can be seen that while $T_{\rm d}$ increases with $L_{\rm IR}$ between $z= 0.45-5.7$, it does not seem to significantly vary at any given redshift. {Right panel}: Since $T_{\rm d}$ variations with $L_{\rm IR}$ at any given redshift remain small, we plot the average peak dust temperature as a function of redshift as black points, with the best-fit solid black line. For comparison, we show the relations of \citet{Schreiber_2018}, \citet{Viero_2022} and \citet{Ismail_2023} as dotted red, dash-dotted green, and dashed blue lines, respectively. For discussion see Sect. \ref{sec:dust}.}
     \label{fig:tz}
\end{figure*}

\subsection{Temperature-redshift relation}\label{sec:dust}

The evolution of the dust temperature with redshift has been debated numerous times in the literature. There are three main sets of claims that can be distinguished -- those that argue that, at fixed $L_{\rm IR}$, $T_{\rm d}$ is lower at higher redshifts (e.g., \citealt{Chapman_2002, Pope_2006, Magnelli_2014}), those that claim $T_{\rm d}$ to be increasing with $z$ (e.g., \citealt{Schreiber_2018, Faisst_2020}), and those that show no evolution at all (e.g., \citealt{Dudzeviciute_2020, Reuter_2020, Drew_2022}). As has been shown by \citet{Drew_2022}, the apparent increase in the dust temperature in lower-redshift galaxies is often caused by the bias, whereby colder systems had been fit with warmer-temperature SEDs due to the lack of long wavelength data (and hence no photometric constraints on the Rayleigh-Jeans tail of the dust emission curve). A number of works that have found no evolution of $T_{\rm d}$ with redshift are based on brightest sources selected at (sub-)mm wavelengths, not necessarily representative of the bulk of the star-forming galaxy population on the MS (e.g., \citealt{Dudzeviciute_2020, Reuter_2020}). \citet{Drew_2022} have found that for the sample of $z<2$ galaxies, the IR luminosity anti-correlates with the dust emission rest-frame peak wavelength (and hence correlates with the luminosity-weighted dust temperature). They have also determined that the $z<2$ $T_{\rm d}$-$L_{\rm IR}$ relation does not evolve with redshift. Given that the SFRs (and $L_{\rm IR}$) for star-forming MS galaxies, at fixed stellar mass, increase with redshift, the redshift-independent correlation of $L_{\rm IR}$ and $T_{\rm d}$ produces hotter SEDs at higher $z$'s.

In the left panel of Fig. \ref{fig:tz} we plot the peak dust temperatures (derived from best-fit SEDs given in Sect. \ref{sec:datav} using Wien's law) for the star-forming sample of galaxies as a function of the IR luminosity, color-coded with redshift. We can see that, as in the case of \citet{Drew_2022}, $T_{\rm d}$ clearly correlates with $L_{\rm IR}$ (even when confining our sample to sources at $z<2$). However, at any given redshift, we do not find any evolution of the dust temperature with $L_{\rm IR}$. If anything, similarly to \citet{Schreiber_2018}, we can see a minor decline of $T_{\rm d}$ at highest IR luminosities at $z\lesssim 2$. As is discussed in \citet{Liang_2019}, the differences in dust temperatures are caused mainly by the variations in the specific star formation rate (sSFR; see the right panel of Fig. \ref{fig:ssfr}). In this scenario, higher sSFR systems have more young ($t\lesssim 10$\,Myrs) star clusters, which heat the dust to higher temperatures, producing more emission in mid-IR wavelengths and hence increasing the values of the effective peak temperatures. This is consistent with the slight decrease in $T_{\rm d}$ at the highest stellar masses and the increase with redshift seen in the left panel of Fig. \ref{fig:tz}. Since, at any given redshift, the dust temperature variations are small, we plot the mean values of $T_{\rm d}$ as a function of $z$ (black dots) in the right panel of Fig. \ref{fig:tz}. We find that the data can be well fit with the quadratic equation, where

\begin{multline}\label{eq:tz}
T^{\rm MS}_{\rm d}/{\rm K}=(1.13\pm 0.17)\times z^2 + (0.88\pm 0.78)\times z\\
+ (21.74\pm 0.70).
\end{multline}

For comparison, the results of \citet{Schreiber_2018}, \citet{Viero_2022} and \citet{Ismail_2023} are shown as dotted red, dash-dotted green, and dashed blue lines, respectively. It can be seen that our results, as well as the relation of \citet{Schreiber_2018}, give slightly higher temperatures at low redshifts than \citet{Ismail_2023}. This is most likely caused by the fact that both in this work and in the work of \citet{Schreiber_2018}, the emissivity index, $\beta$, has been fixed at constant value (1.96 and 1.5, respectively). As is shown in \citet{Ismail_2023}, $\beta$ correlates with dust temperature, where warmer SEDs are characterized by lower emissivity values. At low redshifts, where $T_{\rm d}$ is $\sim 20$\,K, $\beta$ is expected to vary between 2.5 and 3.5. Fixing $\beta$ at lower values will naturally increase the best-fit dust temperatures. The effect is stronger for lower assumed values of emissivity, which can be seen in the figure. Comparing to \citet{Viero_2022}, we can see that our results also point toward slightly hotter dust in the early Universe. As is explained in \citet{Viero_2022}, however, assuming hotter dust at higher redshifts is necessary in order to explain a number of observations, like the lack of dusty objects in the early Universe \citep{Sommovigo_2020}, inconsistencies in the high-$z$ IRX-$\beta$ relation \citep{Bouwens_2016}, or unusually high dust masses \citep{Bakx_2020}.

\subsection{Comparison to the literature} \label{sec:comp}

\begin{figure*}
\centering
   \includegraphics[width=17cm]{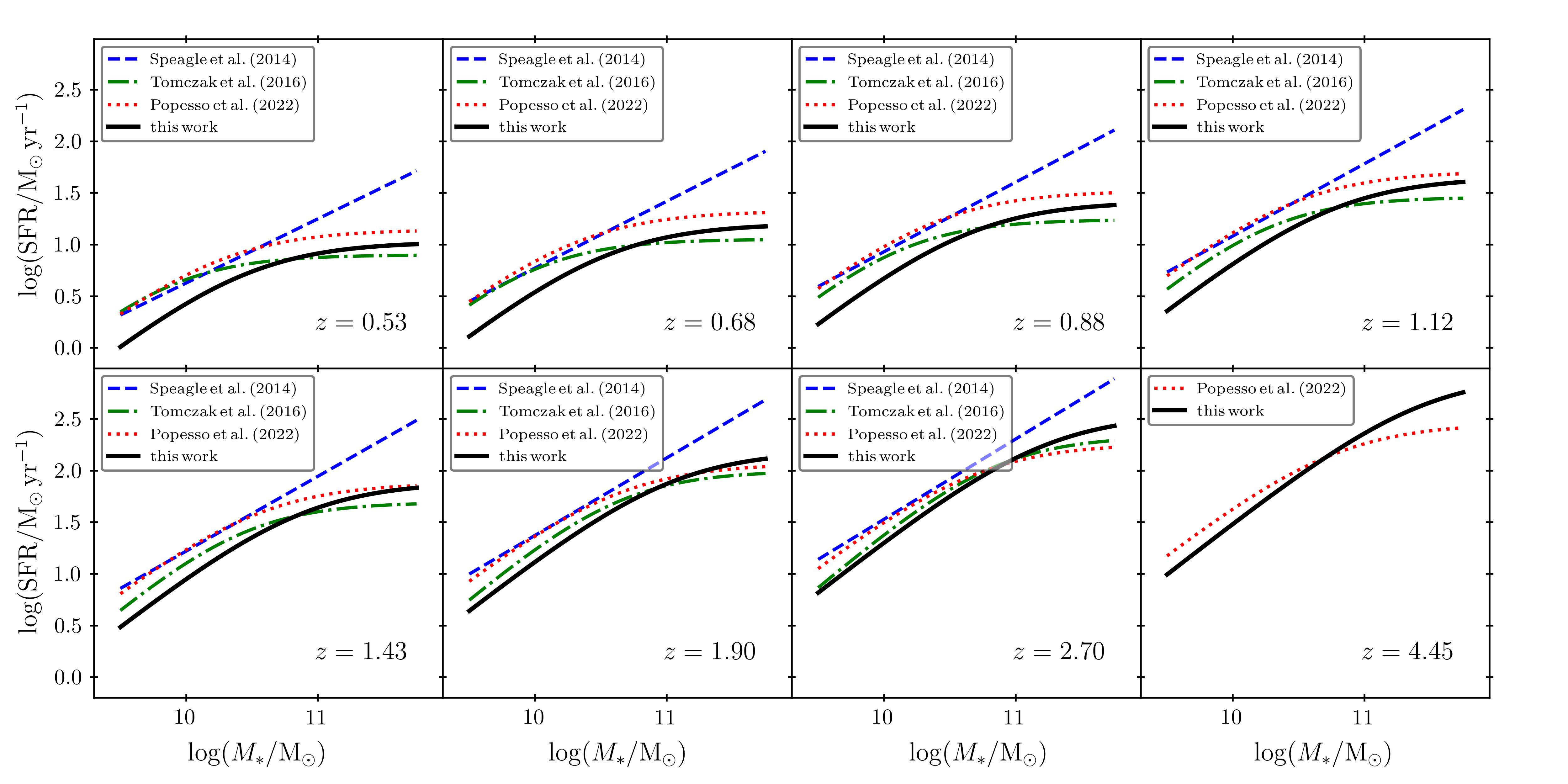}
     \caption{Best-fit star-forming MS shown at each redshift bin studied in this work as the solid black line. For comparison, we show the relations of \citet{Speagle_2014}, \citet{Tomczak_2016} and \citet{Popesso_2023}, plotted with dashed blue, dash-dotted green, and dotted red lines, respectively. While all the functions tend to roughly agree with each other, some inconsistencies can clearly be seen, discussion of which is presented in Sect. \ref{sec:comp}.}
     \label{fig:comp}
\end{figure*}

In Fig. \ref{fig:comp}, we show the comparison of the SFR-$M_\ast$ relation for SFGs (star-forming MS) found in this work (solid black lines) with those found by \citet{Speagle_2014}, \citet{Tomczak_2016} and \citet{Popesso_2023}, plotted with dashed blue, dash-dotted green, and dotted red lines, respectively. The lines represent the best-fit functions at the central value of each redshift bin studied in this work. The first feature that can be seen is the lack of the turnover for the SFR-$M_\ast$ relation of \citet{Speagle_2014}. As has been pointed out by \citet{Popesso_2023}, this was likely caused by the fact that the sample studied in \citet{Speagle_2014} was mostly confined to sources with stellar masses below $10^{11}\,{\rm M_\odot}$. 

At low redshifts, for stellar masses ${\rm log}(M_\ast/{\rm M_\odot})\lesssim 10.5$, our values of the SFR are lower than those of \citet{Tomczak_2016} and \citet{Popesso_2023}. This is most likely caused by fitting our IR data with colder SEDs, where dust temperatures of $\sim 20$\,K were assumed (see Fig. \ref{fig:tz}). At $z>2$ and ${\rm log}(M_\ast/{\rm M_\odot})\gtrsim 11$, our values of SFR tend to be slightly larger than those of \citet{Tomczak_2016} and \citet{Popesso_2023}. As is explained in Sect. \ref{sec:ss}, all the starburst galaxies that were detected with ALMA in COSMOS and UDS fields were excluded from the sample of this work. However, because the relatively long wavelength of ALMA observations tends to miss hot dust sources (e.g., \citealt{Chen_2022}), we likely failed to identify a fraction of a high-mass starburst population that, due to its high SFRs, would boost the average stacked values. In addition,
the determination of the high-mass end of the MS at high redshifts is based on a handful of sources. Looking at Fig.\,3 of \citet{Popesso_2023}, it can be seen that at the highest redshift both the bending and the normalization of the MS are underpredicted by the adopted functional forms.

\begin{figure}
\centering
   \includegraphics[width=7.5cm]{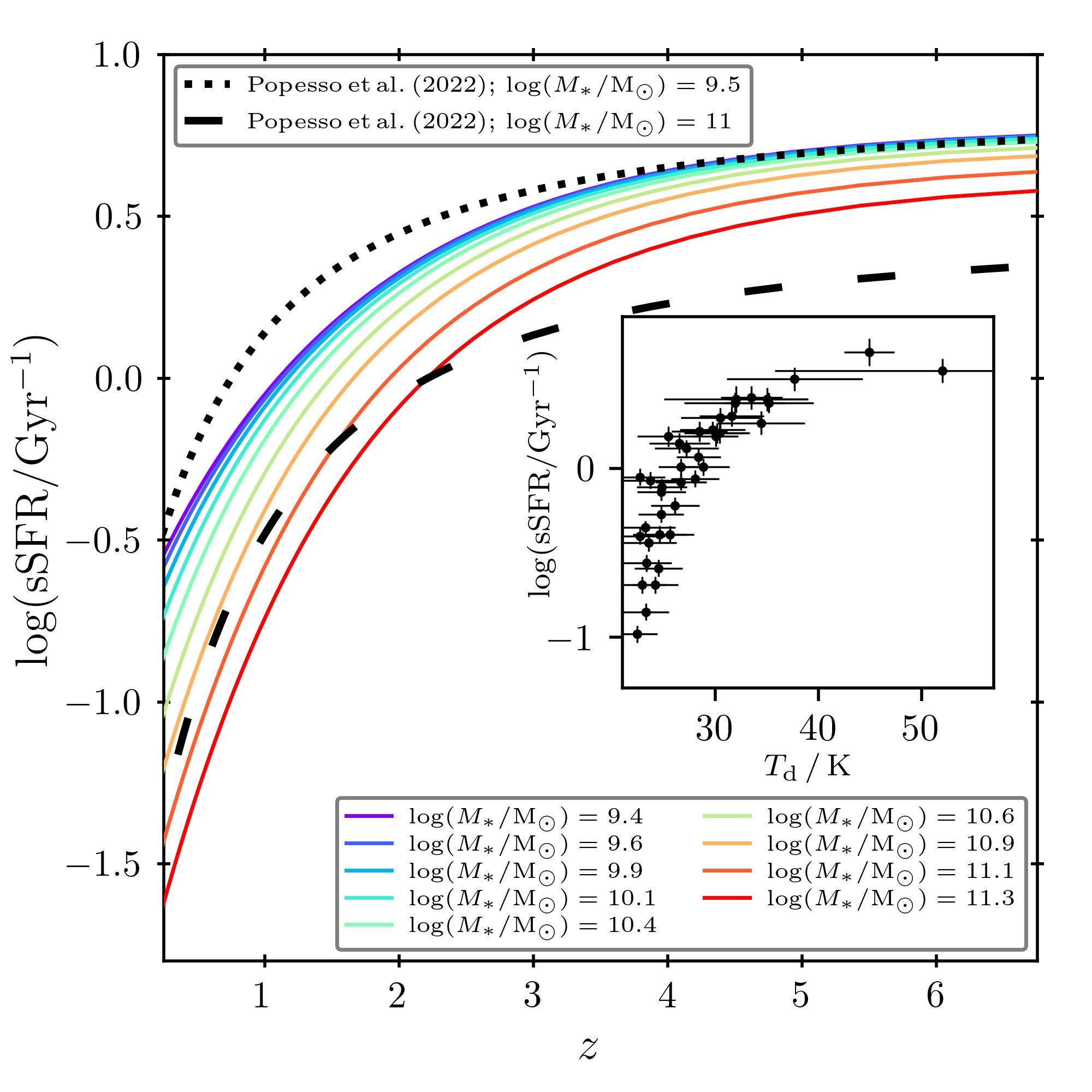}
     \caption{Logarithm of the sSFR, derived from the best-fit functional forms of the star-forming MS found in this work, plotted as a function of redshift and color-coded with stellar mass. The shape of the relations is consistent with the redshift evolution of the molecular gas fraction in SFGs, as is discussed in Sect. \ref{sec:comp}. For comparison, we show the findings of \citet{Popesso_2023} at two stellar mass regimes. The inset plot shown the relation between the sSFR and peak dust temperature. The similar shapes of the sSFR evolution with $z$ and $T_{\rm d}$ is consistent with the scenario, presented in \citet{Liang_2019}, in which different dust temperature are mainly driven by the variations in the sSFR (see Sect. \ref{sec:dust}).}
     \label{fig:ssfr}
\end{figure}

In Fig. \ref{fig:ssfr}, we show the corresponding evolution of the sSFR with redshift, color-coded with the stellar mass. A clear turnover can be seen at $z\sim 2$, which is a direct consequence of the bending of the normalization of the MS with redshift (Eq. \ref{eq:pars}). This, on the other hand, as is explained in Sect. \ref{sec:ev}, is likely caused by the increase in the molecular gas fraction, which was found by \citet{Wang_2022} to be exponentially increasing with redshift. The sSFR-$z$ relation exhibits lower normalization at higher stellar masses, since at masses $\gtrsim M_0$ the MS flattens, decreasing the resulting sSFR. In the inset plot, the evolution of the sSFR with the peak dust temperature can be seen. Similar shapes of the sSFR relation with both redshift and dust temperature are consistent with the suggestion \citep{Liang_2019} that the increase in the dust temperature with redshift, found in Sect. \ref{sec:dust}, is indeed caused by the corresponding increase in the sSFRs. 

For comparison, the relation of \citet{Popesso_2023} for ${\rm log}(M_\ast/{\rm M_\odot})\simeq 10.0$ and 11.5 are shown as dotted and dashed black lines, respectively. The inconsistencies at low and high redshifts come from different normalizations of the corresponding MSs, as was explained above. In order to investigate those further, in the following section we explore how the evolution of the MS found in this work and the works of \citet{Speagle_2014}, \citet{Tomczak_2016} and \citet{Popesso_2023} affects the “growth” of the GSMF.

\subsection{Growth of the galaxy stellar mass function} \label{sec:gsmf}

\begin{figure*}
\centering
   \includegraphics[width=17cm]{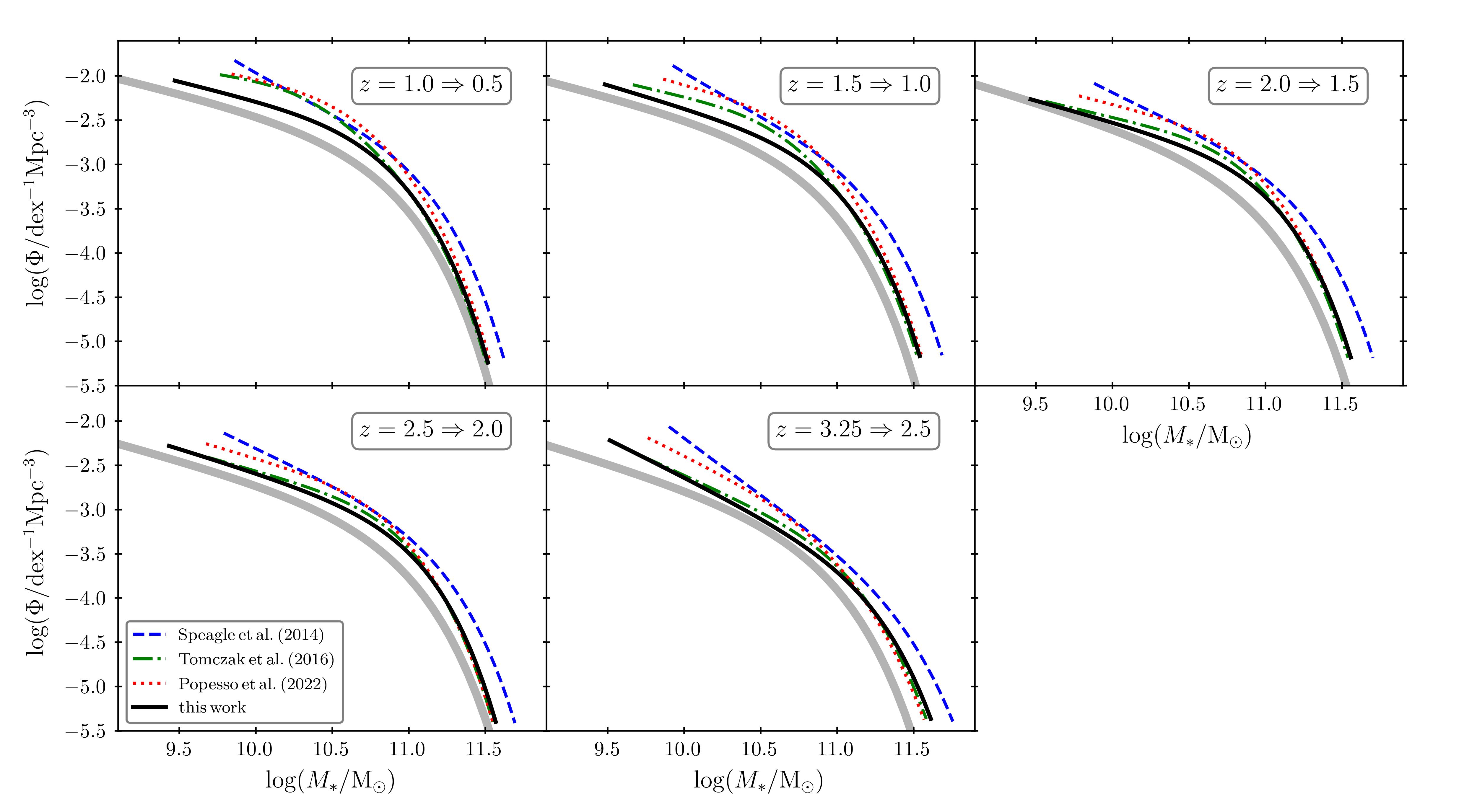}
     \caption{Comparison between the observed and simulated star-forming GSMFs. The simulated GSMFs are constructed by “growing” the observed function of \citet{McLeod_2021}, between the redshifts listed in the upper right corner of each panel, according to Eq. \ref{eq:gsmf}. The mass evolution resulting from assumed star-forming MSs of \citet{Speagle_2014}, \citet{Tomczak_2016} and \citet{Popesso_2023}, are  plotted with dashed blue, dash-dotted green, and dotted red lines, respectively. The simulated function constructed using the results of this work are presented in solid black, while the thick solid gray line represents the observed GSMF at the final redshift of each simulation. The inconsistencies at low- and high-mass ends of the simulated GSMFs come from differences in the corresponding shapes of the MSs involved (Fig. \ref{fig:comp}). The significant shift in densities between the observed and simulated GSMFs is caused by the fraction of SFGs, simulated between the redshifts in question, entering a quiescent mode, as is discussed in Sect. \ref{sec:gsmf}.}
     \label{fig:gsmfsf}
\end{figure*}

\begin{figure*}
\centering
   \includegraphics[width=17cm]{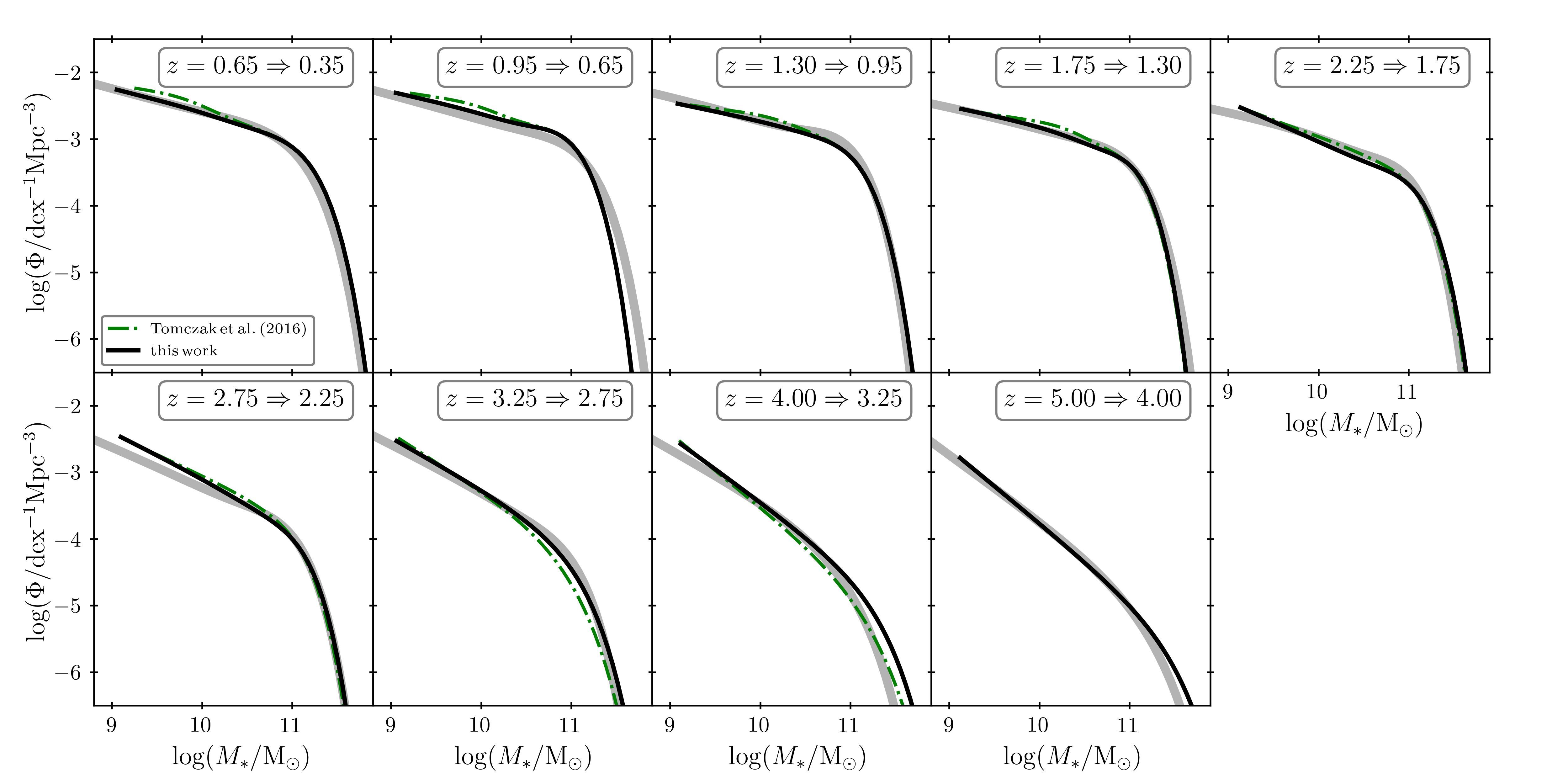}
     \caption{Comparison between the observed and simulated GSMFs, where both star-forming and quiescent galaxies are considered. To extend the redshift range, the simulated GSMFs are constructed by “growing” the observed function of \citet{Davidzon_2017}, between the redshifts listed in the upper right corner of each panel, according to Eq. \ref{eq:gsmf}. The mass evolution results from assuming the SFR-$M_\ast$ relation characteristic of galaxy samples including both the star-forming and quiescent sources, where we compare our results, shown in solid black, with those of \citet{Tomczak_2016}, depicted with a dash-dotted green line. The thick solid gray line represents the observed GSMF at the final redshift of each simulation. It can be seen that the significant gap in densities, apparent in Fig. \ref{fig:gsmfsf}, is no longer present, indicating that, indeed, the inconsistencies were caused by the fraction of sources evolving into quiescent mode, as is discussed in Sect. \ref{sec:gsmf}. The low-mass slope of simulated GSMFs at $z\gtrsim 2$ is inconsistent with the observed function, indicating that the corresponding slope of the SFR-$M_\ast$ relations most likely evolves toward steeper values toward earlier epochs.}
     \label{fig:gsmf}
\end{figure*}

The star-forming MS effectively quantifies the stellar mass growth of a galaxy with time. We can therefore compare the evolution of the GSMF as implied by any given MS with the observed one. The procedure is described in detail in \citet{Leja_2015} and can be summarized as follows. At each redshift, the stellar mass for a given GSMF is being grown assuming the SFR from the MS, causing the GSMF to shift toward higher masses. Additionally, mainly due to winds and outflows, a fraction of the mass is ejected from a stellar population, a careful treatment of which has been presented in \citet{Moster_2013}. The stellar mass growth with time can therefore be expressed as

\begin{equation}\label{eq:gsmf}
\dot{M}(M,z)={\rm SFR}(M,z)\times (1-R),
\end{equation}

\noindent where $R$ is the fraction of the mass lost during passive stellar evolution, starting at 0 and reaching 0.36 after $\sim 10$\,Gyr, assuming the Chabrier IMF. However, since most of the mass loss happens during the first 100\,Myr, following \citet{Leja_2015}, we assume it to be instantaneous and fix $R$ at 0.36. 

We start the procedure with the observed star-forming stellar mass function of \citet{McLeod_2021}. The stellar mass at any given redshift is evolved according to Eq. \ref{eq:gsmf} assuming different star-forming MSs. In Fig. \ref{fig:gsmfsf} we show the results of this exercise, with the observed star-forming GSMF at the final redshift shown in gray. The stellar mass function, produced by evolving the observed GSMF between redshifts listed in the upper right corner of each panel, for the MS of \citet{Speagle_2014}, \citet{Tomczak_2016} and \citet{Popesso_2023} are plotted with dashed blue, dash-dotted green, and dotted red lines, respectively. The results from this work are shown with a solid black line. It can be seen that the evolved GSMFs differ at both low and high-mass ends, which is due to the differences in the shapes of the corresponding MSs. At low redshifts and stellar masses the MS found in this work predicts lower values of the SFR, which causes the GSMF to grow more slowly in mass and hence the shift toward higher masses is in this case smaller than those produced by \citet{Speagle_2014}, \citet{Tomczak_2016} and \citet{Popesso_2023}. Similarly, at high redshift and stellar masses, the stellar mass shifts are slightly larger (except for the case of \citealt{Speagle_2014}, where no bending was assumed). The most striking feature of Fig. \ref{fig:gsmfsf} is the inconsistency of all the evolved functions with the observed GSMF. There are two main reasons for that. Firstly, the mass evolution of the GSMF is going to be affected by galaxy mergers, where mergers are expected to decrease the number density and contribute to the mass growth of galaxies. As is shown by \citet{Leja_2015}, however, including mergers will change the evolved GSMF at low masses by only $0.1-0.2$\,dex between redshifts 2.25 and 0.5. The second reason is the fact that, as GSMF evolves in time, a fraction of SFGs will transform into passive sources, effectively shifting the evolved stellar mass function down toward lower densities.

In order to account for this effect, we additionally evolve the GSMF, where both star-forming and passive galaxies have been included. In order to extend this exercise to higher redshifts, we adopt the stellar mass function of \citet{Davidzon_2017}, where $z\simeq 6$ has been reached. The SFR-$M_\ast$ relation used in this case includes all galaxies (star-forming and quiescent) and is shown in the left panel of Fig. \ref{fig:sfr}. The resulting evolved GSMF at different redshifts is compared to the observed one in Fig. \ref{fig:gsmf} (thin black and thick gray lines, respectively). Since \citet{Speagle_2014} and \citet{Popesso_2023} did not publish the SFR-$M_\ast$ relations for the galaxy samples that include the quiescent sources, we compare our results with \citet{Tomczak_2016} only (dashed green line). It is clear from Fig. \ref{fig:gsmf} that the obvious shift in galaxies number density between the observed and simulated functions, apparent in Fig. \ref{fig:gsmfsf}, is no longer present. This confirms that the inconsistency in the star-forming GSMFs is caused by the decrease in density due to the fraction of SFGs transforming into passive sources. For low masses at $z\gtrsim 2$ the low-mass end of the simulated stellar mass functions tends to be steeper than the observed ones. This is caused by the low-mass slope of the observed GSMF undergoing the evolution in redshift toward higher values. In order for our SFR-$M_\ast$ relation to produce results consistent with this, the low-mass slope, $\gamma$, from Eq. \ref{eq:sfr} would also need to evolve. However, as is explained in Sect. \ref{sec:fit}, since our data does not probe low stellar masses at high redshifts, $\gamma$ was assumed to be constant. The fact that the constant $\gamma$ gives steeper low-mass slopes of the GSMF at high $z$'s indicates that the power law slope of the SFR-$M_\ast$ relation for all galaxies (star-forming and quiescent) should increase in redshift. Similarly, since the low-mass slope of the star-forming GSMF, found in both \citet{Davidzon_2017} and \citet{McLeod_2021}, increases at $z\gtrsim 2$, the slope of the star-forming MS should also evolve. At $z\gtrsim 3$ the high-mass end of the simulated GSMF overpredicts the number of galaxies. While the errors on the observed high-redshift stellar mass function at these masses are significantly large ($\sim 1$\,dex), it is also possible that the high-mass end of the SFR-$M_\ast$ relations in this work, and hence the growth of the stellar masses, is contaminated by the presence of starburst galaxies.

\section{Summary} \label{sec:sum}

We have stacked $\sim 100$k $K$-band selected COSMOS and UDS galaxies in the IR {\it Herschel} and JCMT bands in bins of stellar mass and redshift in order to investigate the SFR evolution out to $z\sim 7$. Inverse variance stacking was used, where, for the {\it Herschel} SPIRE bands, background sources were found to contribute significantly to the stacked fluxes. After removing the background sources' contribution (fitted with the Gaussian function), the IR luminosities, and hence the SFRs, were determined and the functional forms of the SFR evolution with stellar mass and redshift found. This was done for both the star-forming and whole (star-forming and quiescent) galaxy samples, where the quiescent sources were selected using the $UVJ$ color-color space. The main findings of this work can be summarized as follows:

\begin{enumerate}[label=(\roman*),wide]

\item{The SFR of SFGs populating the MS exhibits a nearly linear growth with stellar mass at any given redshift at low masses. Above the so-called bending mass, $M_0$, the MS flattens toward an asymptotic value of the SFR. At constant stellar mass, the SFR also increases with redshift -- the redshift evolution of the normalization of the MS. This behavior is consistent with the scenario in which the evolution of the star-forming MS is driven by the position of the galaxy on the integrated SK relation, given its molecular mass fraction (e.g., \citealt{Santini_2014, Wang_2022}). We find that the redshift evolution of the bending mass, $M_0$, is exponential, increasing sharply from $z\sim 0.5$ to $z\sim 2$ and flattening at $z>2$. The similarity to the shape of the relation between the stellar mass, above which shocks can efficiently heat incoming baryons, and redshift, found in \citet{Daddi_2022}, suggests that the flattening of the MS is likely caused by the phasing out of the cold accretion.}

\item{We investigated the evolution of the peak dust temperature with IR luminosity and redshift, by fitting the stacked photometry with the dust emission curve of 
\citet{Casey_2012}. Similarly to \citet{Drew_2022}, an increase in $T_{\rm d}$ with IR luminosity was found. However, when confined to any given redshift, our data does not show any significant evolution with $L_{\rm IR}$. We conclude that the increase in $T_{\rm d}$ with IR luminosity is driven by the evolution of the $L_{\rm IR}$ with redshift for the MS galaxies. We find that the dust temperature increases quadratically with redshift, consistent with the scenario in which higher values of $T_{\rm d}$ for SFGs are driven by the increase in their sSFRs (e.g., \citealt{Liang_2019}).}

\item{The evolution of the MS found in this work was compared with the relations of \citet{Speagle_2014}, \citet{Tomczak_2016} and \citet{Popesso_2023}. We find that the normalization of our MS at $z\lesssim 2$ for masses below the bending mass, $M_0$, is slightly lower, while the SFRs above $M_0$ at high redshifts exhibit larger values (except for the MS of \citealt{Speagle_2014}, where no MS bending was found). We attribute these inconsistencies to the dust temperatures assumed, when fitting the IR data. As was found in this work, $T_{\rm d}$ increases from $\sim 20$\,K to $\sim 60$\,K between redshifts 0.5 and 5.5, and hence assuming dust temperatures at the level of $\sim 30$\,K, as is often done in the absence of the FIR data, will correspondingly affect the determined IR luminosities (and hence the SFRs). We also note that the higher values of the SFRs at high redshifts for $M_\ast>10^{11}\,{\rm M_\odot}$ could be, at least in part, caused by the contamination of the MS sample by starburst galaxies.}

\item{Adopting different shapes of the SFR-$M_\ast$ relation, we simulated the time evolution for the observed GSMFs of \citet{Davidzon_2017} and \citet{McLeod_2021}. For the star-forming sample, we find that the GSMF “grows” too fast, producing simulated relations significantly above the observed ones. By performing the simulations using the SFR-$M_\ast$ relation, found using both star-forming and quiescent galaxies, we confirm that the overprediction of SFG number densities is caused by a significant fraction of sources evolving into the quiescent phase and thereby leaving the star-forming MS. In addition, we find the simulated low-mass end of the GSMF gives higher number densities at high $z$, indicating that the low-mass slope of the MS likely evolves toward steeper values with redshift. However, due to the significant uncertainties, more detailed analysis based on larger high-$z$ samples is required.}

\end{enumerate}

\section{Data availability}
\label{sec:datav}

Example stacked image and best-fit SEDs are available \href{https://zenodo.org/records/13843224?token=eyJhbGciOiJIUzUxMiJ9.eyJpZCI6IjM5MTJhNGFiLTMxNjAtNGI5OC04NDk3LTgzZjY4OWY4N2QxYSIsImRhdGEiOnt9LCJyYW5kb20iOiIwZWVkODdlZWEwOTljNmM3ZDBjNWI2ZDI3ZGM5Yzc2MiJ9.IKzxuAo4NFyL_kk4in-Za0FkyX0gTa1CXqRNn0RlPcVFKjtGKBgVT1s-6WV4rptnaupnLXWo5TIpjefG1oh0aA}{here}.

\begin{acknowledgements}

This research was funded in whole or in part by the National Science Center, Poland (grant no. 2020/39/D/ST9/03078). For the purpose of Open Access, the author has applied a CC-BY public copyright license to any Author Accepted Manuscript (AAM) version arising from this submission. JSD acknowledges the support of the Royal Society through a Research Professorship. MJM acknowledges the support of the National Science Centre, Poland through the SONATA BIS grant 2018/30/E/ST9/00208 and the Polish National Agency for Academic Exchange Bekker grant BPN/BEK/2022/1/00110. DJM acknowledges the support of the Science and Technology Facilities Council. KL acknowledges the support of the Polish Ministry of Education and Science through the grant PN/01/0034/2022 under `Perły Nauki' program.

\end{acknowledgements}

\bibliographystyle{aa}    
\bibliography{papers}

\begin{appendix}

\section{Additional table}\label{sec:ap1}

\begin{sidewaystable*}
\tiny
\caption{Stacked values of the SFR (in $\Msun\,{\rm yr}^{-1}$) for all the galaxies (star-forming and quiescent; top panel) and SFGs (bottom panel) between redshifts 0.45 and 5.7. The first column lists the mass bins, where $\mathcal{M}~\equiv~\log_{10}(M_{\star}/\Msun)$.}\label{tab:sfr}
\begin{tabular}{ccccccccc}
% \tablewidth{700pt}
% \tabletypesize{\scriptsize}
\hline\hline
& $0.45\leq z<0.60$ & $0.60\leq z<0.75$ & $0.75\leq z<1.00$ & $1.00\leq z<1.25$ & $1.25\leq z<1.60$ & $1.60\leq z<2.20$ & $2.20\leq z<3.20$ & $3.20\leq z<5.70$ \\ 
& log(SFR) & log(SFR) & log(SFR) & log(SFR) & log(SFR) & log(SFR) & log(SFR) & log(SFR) \\
\hline
$\phantom{0}9.25\leq \mathcal{M} < \phantom{0}9.50$ & $0.01\pm 0.06$ & $0.20\pm 0.05$ & $0.28\pm 0.06$ & $0.51\pm 0.06$ & & & & \\ 
$\phantom{0}9.50\leq \mathcal{M} < \phantom{0}9.75$ & $0.18\pm 0.05$ & $0.42\pm 0.05$ & $0.51\pm 0.05$ & $0.68\pm 0.05$ & $0.85\pm 0.05$ & & & \\ 
$\phantom{0}9.75\leq \mathcal{M} < 10.00$ & $0.39\pm 0.05$ & $0.54\pm 0.05$ & $0.66\pm 0.05$ & $0.84\pm 0.05$ & $1.04\pm 0.06$ & $1.21\pm 0.06$ & & \\ 
$10.00\leq \mathcal{M} < 10.25$ & $0.44\pm 0.05$ & $0.65\pm 0.05$ & $0.81\pm 0.05$ & $0.97\pm 0.05$ & $1.18\pm 0.05$ & $1.33\pm 0.06$ & $1.57\pm 0.06$ & \\ 
$10.25\leq \mathcal{M} < 10.50$ & $0.55\pm 0.05$ & $0.64\pm 0.06$ & $0.86\pm 0.05$ & $1.11\pm 0.05$ & $1.28\pm 0.05$ & $1.49\pm 0.05$ & $1.68\pm 0.06$ & $2.10\pm 0.08$ \\ 
$10.50\leq \mathcal{M} < 10.75$ & $0.47\pm 0.06$ & $0.74\pm 0.05$ & $0.90\pm 0.05$ & $1.17\pm 0.05$ & $1.41\pm 0.05$ & $1.54\pm 0.05$ & $1.80\pm 0.06$ & $2.22\pm 0.09$ \\ 
$10.75\leq \mathcal{M} < 11.00$ & $0.70\pm 0.06$ & $0.78\pm 0.06$ & $0.85\pm 0.06$ & $1.26\pm 0.05$ & $1.44\pm 0.05$ & $1.73\pm 0.05$ & $2.01\pm 0.06$ & $2.38\pm 0.09$ \\ 
$11.00\leq \mathcal{M} < 11.25$ & $0.61\pm 0.08$ & $0.79\pm 0.07$ & $0.92\pm 0.06$ & $1.27\pm 0.06$ & $1.55\pm 0.06$ & $1.81\pm 0.06$ & $2.20\pm 0.06$ & $2.60\pm 0.08$ \\ 
$11.25\leq \mathcal{M} < 11.50$ & -- & $1.02\pm 0.08$ & -- & $1.53\pm 0.08$ & -- & $2.06\pm 0.07$ & $2.31\pm 0.09$ & $2.71\pm 0.10$ \\ 
\hline
$\phantom{0}9.25\leq \mathcal{M} < \phantom{0}9.50$ & $0.05\pm 0.05$ & $0.26\pm 0.06$ & $0.35\pm 0.07$ & $0.52\pm 0.06$ & & & & \\ 
$\phantom{0}9.50\leq \mathcal{M} < \phantom{0}9.75$ & $0.25\pm 0.05$ & $0.49\pm 0.05$ & $0.57\pm 0.05$ & $0.74\pm 0.05$ & $0.87\pm 0.06$ & & & \\ 
$\phantom{0}9.75\leq \mathcal{M} < 10.00$ & $0.50\pm 0.05$ & $0.69\pm 0.06$ & $0.81\pm 0.06$ & $0.91\pm 0.06$ & $1.12\pm 0.06$ & $1.24\pm 0.06$ & & \\ 
$10.00\leq \mathcal{M} < 10.25$ & $0.62\pm 0.06$ & $0.84\pm 0.06$ & $1.00\pm 0.05$ & $1.10\pm 0.05$ & $1.28\pm 0.06$ & $1.43\pm 0.06$ & $1.61\pm 0.06$ & \\ 
$10.25\leq \mathcal{M} < 10.50$ & $0.76\pm 0.05$ & $0.92\pm 0.06$ & $1.12\pm 0.05$ & $1.28\pm 0.05$ & $1.44\pm 0.05$ & $1.62\pm 0.05$ & $1.73\pm 0.06$ & $2.16\pm 0.08$ \\ 
$10.50\leq \mathcal{M} < 10.75$ & $0.77\pm 0.06$ & $1.06\pm 0.06$ & $1.24\pm 0.05$ & $1.39\pm 0.05$ & $1.57\pm 0.05$ & $1.70\pm 0.05$ & $1.90\pm 0.06$ & $2.26\pm 0.10$ \\ 
$10.75\leq \mathcal{M} < 11.00$ & $1.01\pm 0.06$ & $1.11\pm 0.06$ & $1.25\pm 0.06$ & $1.49\pm 0.06$ & $1.63\pm 0.06$ & $1.88\pm 0.06$ & $2.08\pm 0.06$ & $2.38\pm 0.09$ \\ 
$11.00\leq \mathcal{M} < 11.25$ & $0.96\pm 0.11$ & $1.17\pm 0.09$ & $1.42\pm 0.08$ & $1.52\pm 0.07$ & $1.79\pm 0.07$ & $1.96\pm 0.06$ & $2.28\pm 0.06$ & $2.62\pm 0.08$ \\ 
$11.25\leq \mathcal{M} < 11.50$ & -- & $1.14\pm 0.32$ & $1.39\pm 0.14$ & $1.73\pm 0.10$ & -- & $2.27\pm 0.08$ & $2.42\pm 0.10$ & $2.74\pm 0.09$ \\ 
\end{tabular}
\end{sidewaystable*}

\end{appendix}

\end{document}